\documentclass[11pt]{article}%
\usepackage{amsmath}
\usepackage{amsfonts}
\usepackage{amssymb}
\usepackage{graphicx}%
\begin{document}
% This is the arXiv version of the SITR paper
% submission nlin.drhsedaghat.22311 3/18/2008

\centerline{{\large \textbf{\hspace*{.25in} Complex patterns of spontaneous initiation and termination}}}
\centerline{{\large \textbf{of reentrant circulation in a loop of cardiac tissue}}\footnotetext{This research was supported in part by a grant from Medtronic, Inc.}}

\vspace{2ex}

\centerline{ H. Sedaghat$^1$\dag \footnotetext{\dag Corresponding author; Email: hsedagha@vcu.edu}, 
M.A. Wood$^2$ , J.W. Cain$^1$, C.K. Cheng$^3$ }
\centerline{ C.M. Baumgarten$^4$ , D.M. Chan$^1$ }

\vspace{2ex}

\centerline{{\small $^1$Department of Mathematics and the Center for the Study of Biological Complexity;}} 
\centerline{{\small $^2$Department of Internal Medicine-Cardiology and the Pauley Heart Center;}}
\centerline{{\small $^3$Department of Computer Science and the Center for the Study of Biological Complexity;}}
\centerline{{\small $^4$Department of Physiology and the Pauley Heart Center;}}
\centerline{{\small Virginia Commonwealth University, Richmond, Virginia, 23284-2014, USA}}

\vspace{2ex}

\begin{abstract}
\noindent A two-component model is developed consisting of a discrete loop of
cardiac cells that circulates action potentials as well as a pacing mechanism. 
Physiological properties of cells such as restitutions of refractoriness
and of conduction velocity are given via experimentally measured functions.
The dynamics of circulating pulses and their interactions with the pacer are
regulated by two threshold relations. Patterns of spontaneous initiations and
terminations of reentry (SITR) generated by this system are studied through
numerical simulations and analytical observations. These patterns can be
regular or irregular; causes of irregularities are identified as the threshold
bistability of reentrant circulation (T-bistability) and in some cases, also 
phase-resetting interactions with the pacer.
\end{abstract}


\vspace{2ex}

\section{Introduction.}

Ventricular arrhythmia is the leading cause of cardiac arrest and sudden
death. Clinical observations and implantable cardioverter defibrillators have
accumulated a substantial amount of data on the occurrences of ventricular
arrhythmia in patients [1, 36, 37, 38, 52, 59-61]. Temporal patterns of
initiations and terminations of ventricular arrhythmia tend to exhibit substantial
variations across different time scales, and their occurrences do not correlate
decisively with medication, exertion, stress, lifestyles and similar factors.
Arrhythmia events are not random; they show circadian patterns [36, 60] and
also tend to occur in clusters. However, the detection times between
consecutive events and clusters are spread out over time [37, 52, 59], making
it difficult to understand their causes and make predictions about their
occurrences. Unlike the circadian patterns, these clusterings or their
patterns of occurrences are not affected by the long-term administration of
antiarrhythmic drugs [59]. In spite of the abundant data in existence, the
basis for non-circadian patterns is not well-understood.

Many factors, ranging from internal cardiac mechanisms to external chance
events play significant roles in shaping the electrocardiogram
(ECG)\ recordings and the implantable defibrillator data. Separating all of
the possible contributions is a formidable task, but understanding the
influences of various factors and the extent to which each plays a role may
have important consequences for the diagnosis and treatment of
tachyarrhythmias. A number of prior studies using special preparations of
animal cell cultures establish that complex, spontaneously generated patterns
may occur without some of the features peculiar to the heart (e.g. the
3-dimensional geometry or biological features such as valves, different tissue
layers and types, etc) [3, 25, 35, 44].\ These studies also indicate (both
theoretically and experimentally) that causes external to the heart, whether
random or deterministic, are not always necessary for the occurrence of
irregular behavior. A different class of studies that involve interactive
self-oscillatory sources also establish the capability of basic cardiac
mechanisms to generate complex rhythms without considering the whole heart
[23, 28, 43].

From these and similar studies it may be inferred that changes in the basic
internal cardiac mechanisms can play a significant role in generating and
sustaining arrhythmias. To clarify the role of these basic internal mechanisms
at a fundamental level, in this paper we focus on the unidirectional reentrant
circulation of action potentials in a loop of ventricular tissue. Because of
their pervasiveness, relative simplicity and importance in clinical arrhythmogenesis
reentrant loops have received a great deal of attention [4, 8, 11, 14-16, 20, 26, 30,
41, 44, 48-50, 55]. Studying spontaneous initiation and termination of reentry
(SITR) patterns in the loop context is considerably simpler than in the whole
heart and can offer potentially useful insights into the clinically observed
patterns of arrhythmia. In the construction of a workable model, a certain
level of abstraction bridges over physiological complexities that do not play
a central role in the long-term evolution of temporal patterns.

In this paper we consider a model that combines the traditional discrete loop
with a pacer to form an interactive system. Fundamental thresholds are
dynamically integrated so they can either trigger reentry or inhibit it.
Although a relatively simple mechanism, this composite system has two
important features: It is an essential component of a well-known cardiac
anomaly; and it is capable of generating complex initiation and termination patterns.

In a series of case-studies we show that the loop-pacer mechanism may be
responsible for irregularities in onset and termination of reentry beyond such
pervasive factors as tissue heterogeneity, the geometry of the heart and other
physiological and non-physiological factors. Even without such complexities,
the SITR\ patterns generated by the loop-pacer system can be complex; i.e.
they contain many regular features yet their evolution over time is difficult
to predict. We trace this complexity to the interaction of a threshold with
bistability of the reentrant circulation and in some cases, also to the
inherent discontinuity of phase resetting interactions between the loop and
the pacer. Bistability is an inherent physiological characteristic that may be
attributed to non-uniformities in the\ restitution of action potential
duration (APD); this feature emerges prominently when conduction velocity (CV)
is properly matched with APD. When this happens (e.g. when the length of the
loop is within a certain range) the SITR patterns become considerably more
complex and unpredictable.

\section{The threshold model.}

In this section we describe the main features of the loop-pacer model, leaving
a few additional details to Section 4. The pacer is a self-oscillatory
mechanism that is tied to the reentrant circuit (the loop in this paper) via a
time lag or delay parameter and a special threshold.

\subsection{The loop.}

Consider a loop of cardiac tissue consisting of cells that conduct action
potentials and assume that a single conducting pathway connects the loop to
the rest of the heart. Let the physical length of the loop be denoted by $L$
measured in centimeters. Divide the loop into $m$ sets or aggregates of cells,
each of which may be called a \textit{cell aggregate} or for brevity, just a
\textquotedblleft cell.\textquotedblright\ The cell aggregate that is
connected to the only pathway out of the loop is also the gateway through
which pulses enter the loop or exit it; we label it Cell 1; see Figure 1. This
cell is adjacent not only to Cell 2, but also to Cell $m$ at its other end.

\begin{figure}[ptb]
\label{schematicloop}
\par
\begin{center}
\includegraphics[width=2.0in]{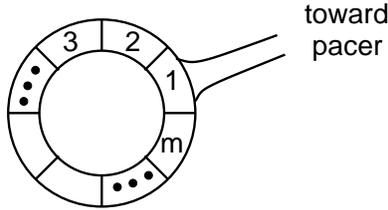}
\end{center}
\caption{\small Schematic diagram of a loop joined to the rest of the heart by a
single conducting pathway.}%
\end{figure}

For each integer $i$ between $1$ and $m,$ label the length of the $i$-th Cell
$\Delta L_{i}.$ These cell aggregates' lengths are not necessarily equal but
of course, they add up to $L.$ Since each cell aggregate must have at least
one cardiac cell in it, $\Delta L_{i}$ has a natural lower bound, namely the
nominal length of a single cardiac cell (about 0.01 cm or 100 microns). The closer all
$\Delta L_{i}$ are to this lower bound, the greater is the number $m$ of cell
aggregates in a loop of fixed length $L$; for a 12 cm loop with cells of
length 0.01 cm, the number of cells is 1200. While for discrete modeling it is
not necessary to pick $m$ that large, it needs to be large enough (hence each
cell aggregate small enough) that the conduction velocity from one end of an
aggregate to the other can be taken to be approximately constant. This
assumption is essential if the restitution of conduction time (CT) is defined
in terms of the restitution of CV, as we do later on.

We assume that all the cardiac cells within a given aggregate are identical.
If all cells in the \textit{entire loop} have identical physiological
characteristics then we may also let cell aggregates all have equal lengths.
In such a case, the loop is said to be \textquotedblleft
homogeneous.\textquotedblright\ Otherwise the loop is \textquotedblleft
heterogeneous.\textquotedblright

To initiate unidirectional circulation, it is necessary that a unidirectional
block (UB) exist somewhere in the loop; for simplicity, we place it in Cell 1,
the gateway to the loop; see Figure 1. Thus Cell 1 is able to activate Cell 2
but not Cell $m.$ For our simulations, we use a simplified version of the UB time
window that is defined by a threshold value for the diastolic interval (DI). Thus conduction is blocked in any cell whose DI is not greater than the threshold 
$DI^{\ast}$; see Section 4 below for more details.
Such a minimum value is used in [26] as a termination mechanism for explaining
the experimental results of [20]; also see [19].

\subsection{Restitution functions.}

We use the term \textquotedblleft restitution\textquotedblright\ generally for
relations that give a particular quantity, e.g. action potential duration or
conduction time, as a function of the diastolic interval (DI). The DI (usually
measured in milliseconds, ms) is the rest or recovery period for the cell
which essentially starts with the end of the refractory period and ends when
the cell is activated again.

\subsubsection{\textbf{APD restitution.}}

In its most basic form, the action potential duration or APD is the length of
time (usually measured in milliseconds, ms) that a cell is active after
excitation. For our purposes, we may think of APD as a cell's effective
refractory period (ERP) during which no excitations, even strong ones, can
elicit new action potentials. The restitution of APD, which can be
experimentally measured or derived from ionic models, is the most extensively
studied of restitution relations [2, 4-8, 10, 11, 21, 24, 26, 29, 32, 33, 42, 45,
48, 50, 53-57]. Any APD restitution function, whether experimentally measured
or analytically derived can be used in this model.\ For the numerical
simulations in this paper, we use an APD\ restitution function that is fitted
to the experimental data by Koller, et al [33] where APD values were recorded
in two types of patients, those with and those without structural heart
disease (SHD). The following is a possible fit to their averaged data
displayed in Figure 1 in [33] for SHD\ patients:%

\begin{equation}
A(DI)=a_{1}-a_{2}e^{-\sigma_{1}DI}-a_{3}e^{-\sigma_{2}(DI-\tau_{1})}%
-a_{4}e^{-\sigma_{3}(DI-\tau_{2})^{2}}+\frac{a_{5}(DI-\tau_{3})}{(DI-\tau
_{3})^{2}+a_{6}} \label{apd-shd}%
\end{equation}
with parameter values

\begin{center}%
\begin{tabular}
[c]{|c|c|c|c|c|c|c|c|c|c|c|c|}\hline
$a_{1}$ & $a_{2}$ & $a_{3}$ & $a_{4}$ & $a_{5}$ & $a_{6}$ & $\sigma_{1}$ &
$\sigma_{2}$ & $\sigma_{3}$ & $\tau_{1}$ & $\tau_{2}$ & $\tau_{3}%
$\\\hline\hline
350 & 157 & 8 & 20 & 1700 & 1200 & 0.0021 & 0.025 & 0.0004 & 80 & 136 &
82\\\hline
\end{tabular}

\end{center}

A graph of this function is shown in Figure 2. Expressing the APD as a function of DI in the above manner represents only a first order of approximation (see Section 4 below).

\begin{figure}[ptb]
\label{apdrc}
\par
\begin{center}
\includegraphics[width=3.0in]{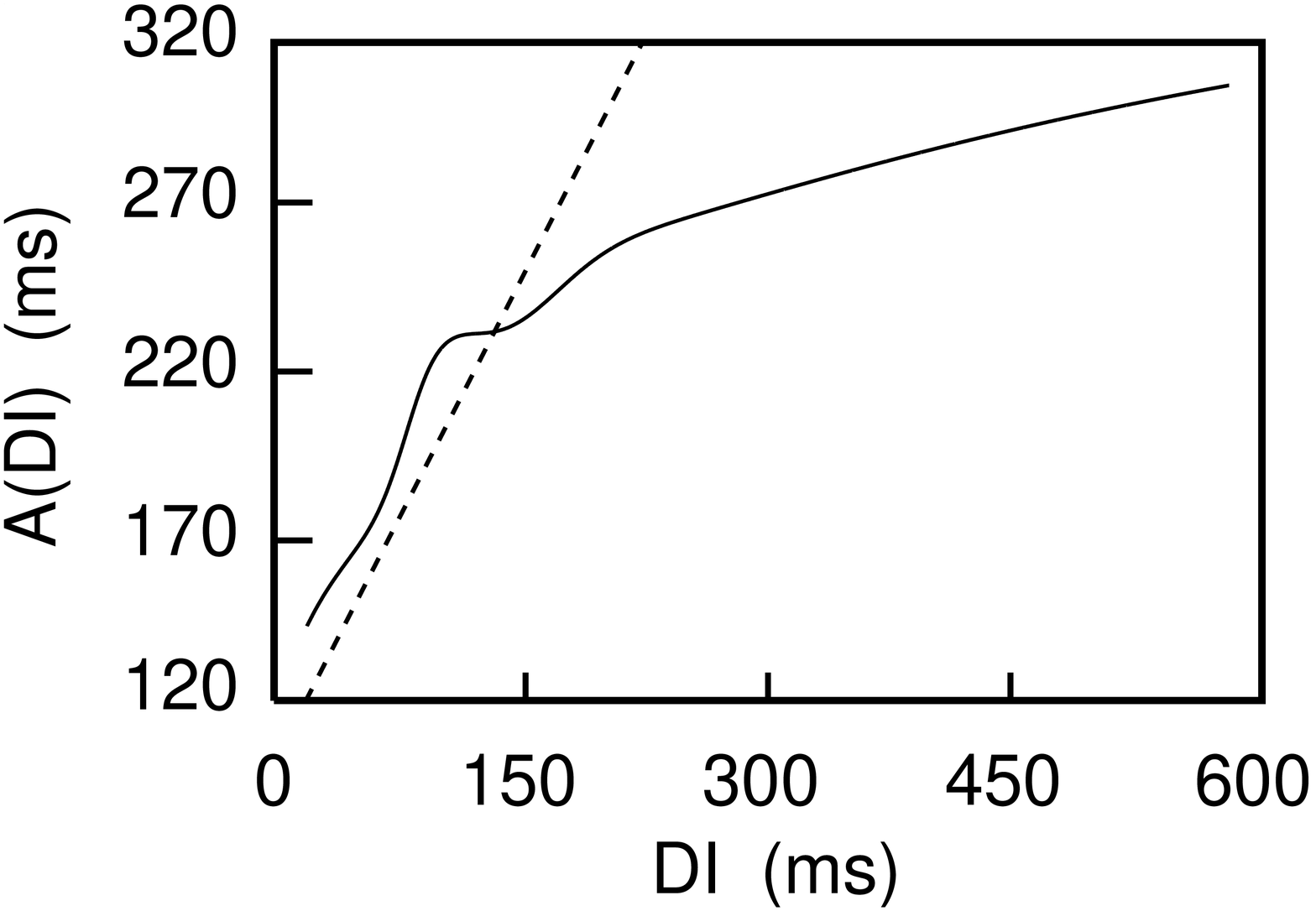}
\end{center}
\caption{\small The APD restitution curve given by~\eqref{apd-shd}. The
dashed line has slope 1 and is included for reference.}%
\end{figure}\medskip

\subsubsection{\noindent\textbf{CT restitution. }}

Each cell conducts an action potential through it in a finite amount of time.
This time interval, usually measured in milliseconds, is referred to as the
conduction time (CT) [4, 26, 50]. Experimentally measured CT restitution
functions are not as readily available for human hearts as the APD restitution
functions; however, they may be readily derived from CV restitution functions
through the relation%
\begin{equation}
C(DI)=\frac{\Delta L}{V(DI)} \label{CV2CT}%
\end{equation}
where $C$ and $V$ are, respectively, the CT and the CV restitution functions.
This derivation is valid if as mentioned earlier, $V$ is essentially constant
over the length of each cell (or cell aggregate) and thus no spatial
dependence is required in $V.$

For discussions of CV restitution functions see [2, 7, 8, 11, 12, 14-16, 20, 22, 53,
54, 56]. For our simulations we use the following expression
\begin{equation}
C(DI)=\frac{\Delta L}{c}[1+de^{-\omega DI}],\quad DI>0 \label{CT}%
\end{equation}
with parameter values:

\begin{center}%
\begin{tabular}
[c]{|c|c|c|c|}\hline
$\Delta L$ & $c$ & $d$ & $\omega$\\\hline\hline
0.1 cm & 0.07 cm/ms & 1 & 0.02 ms$^{-1}$\\\hline
\end{tabular}

\end{center}

These numbers are obtained using (\ref{CV2CT}) and a reasonable fit to the
human-rescaled, guinea pig data for CV from [22, 54]. The value of $\Delta L$
corresponds to 10 nominal cells. The number $c$ in (\ref{CT})is the 
maximum CV; in more common (though less uniform) units, the fitted number in the 
above table would be 0.7 m/s. By changing the parameters in (\ref{CT}) we may
achieve conduction slow-downs in different ways, as shown in Figure 3. 

\begin{figure}[ptb]
\label{ctrc}
\par
\begin{center}
\includegraphics[width=3.0in]{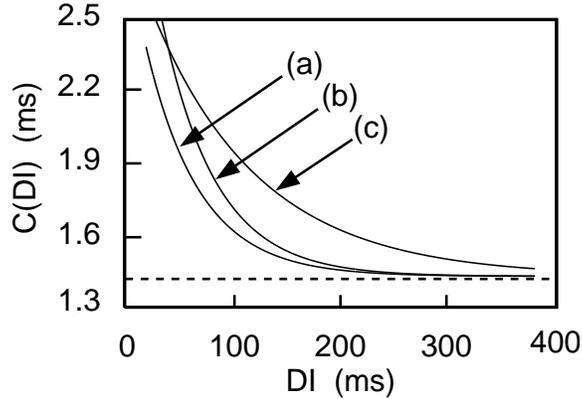}
\end{center}
\caption{\small CT restitution curves given by~\eqref{CT} for several
choices of parameters. For each curve, $\Delta L = 0.1$ and $c = 0.07$. (a) $d
= 1$ and $\omega= 0.02$; (b) $d = 1.5$ and $\omega= 0.02$; and (c) $d = 1$ and
$\omega= 0.01$.}%
\end{figure}

\subsection{\textbf{The two rhythms}.}

In the absence of reentry, a pacer (the sinus node or an ectopic focus) drives
the ventricular contraction cycle. If reentry is initiated and interrupted then the pacer and the loop fall out of phase and it is necessary to carefully model their
interaction. For this purpose we define two rhythms or clocks: The pacer's
rhythm which has a prescribed beat pattern (fixed or variable) and the system's rhythm which is reset when reentry is initiated or terminated. These two rhythms are related via a pacer threshold (seen the next section) and a delay term that we now define.

The \textit{time of occurrence} for the $n$-th beat (following a fixed
reference beat) can be tracked and calculated using the following quantity,
namely, the sum of usually variable cycle lengths (after a DI adjustment):

\[
\rho_{n}=\sum_{j=1}^{n}CL_{j}+(DI_{1,0}-DI_{1,n}),\quad CL_{j}=A(DI_{1,\,j-1}%
)+DI_{1,\,j}.
\]

Let the time of occurrence of the $k$-th pacer beat be denoted by $\beta_{k}$
(starting from the same reference point in time as that for $\rho_{n}$). The
difference
\[
\beta_{k}-\beta_{k-1}=B(k)
\]
is the duration of each such beat. In our simulations $B(k)$ is defined
arbitrarily in order to facilitate our study of the SITR patterns. If the
pacer is the sinus node then $B(k)$ is influenced by a variety of
deterministic and stochastic factors, such as circadian, weekly and seasonal
variations, lifestyles, drugs, the autonomic system and so on. Thus a single,
all purpose formula for $B(k)$ does not exist.

The two rhythms $\rho_{n}$ and $\beta_{k}$ fall out of phase if reentry occurs
as noted earlier. To model their interaction, we define the \textit{first}
pacer pulse that reaches Cell 1 in beat $n$ as the\textit{ least }integer
$k_{n}$ such that
\begin{equation}
\beta_{k_{n}}>\rho_{n}+\delta_{n}.\label{kn}%
\end{equation}

The variable delay term $\delta_{n}$ here represents a\textit{ time lag }that
results from the blocking of the pacer by the retrograde reentrant wave propagating
toward the pacer. A number of things may affect $\delta_{n}$; these include the different types of tissue in which pulses propagate (retrograde and antegrade), the electrotonic spread of current in conducting tissue between the pacer and the loop, and the effect of reentrant waves on the pacer's intrinsic beat rate. For simplicity we do not consider the electrotonic currents and tissue heterogeneity. We assume also that the pacer itself is protected in the sense that it's intrinsic beat rate is not affected by the reentrant activity.

Therefore, we use a simple definition for $\delta_{n}$ as follows:%
\[
\delta_{n}=\left\{
\begin{array}
[c]{ll}%
\delta, & \text{if pulse }n-1\text{ was reentrant}\\
DI^{\ast}, & \text{otherwise }%
\end{array}
\right.  ,\ \delta>DI^{\ast}.
\]

If the loop is not in\ reentry, then $\delta_{n}$ may take the minimum value
$DI^{\ast}$ since Cell 1 must have at least that much rest time after its ERP
before it can be reactivated. High frequency reentrant waves typically
disassociate a slow pacer such as the sinus node. This situation is modeled
here by choosing a sufficiently large value of the delay parameter $\delta$ so
as to inhibit the pacer from interfering with the reentrant circulation. Small
values for $\delta$, which promote pacer interference, may be feasible in
certain circumstances, e.g. when there are pathways from the pacer's site to
the site of the loop that are protected from the fast reentrant waves by pathologies.

To determine the index value $k_{n}$ we may generate the sequence $\beta_{k}$
independently of $\rho_{n}$ and for each $n$ mark the values $k_{n}$ that
satisfy (\ref{kn}). This process is carried out internally during each
simulation run; an analytical relationship is generally not easy to find.
However, if $B(k)=B_{0}$ is constant then such a relationship is easy to find;
in this case we have $\beta_{k_{n}}=k_{n}B_{0}$ with
\[
k_{n}=\left\lceil \frac{\rho_{n}+\delta_{n}}{B_{0}}\right\rceil
\]
i.e., the least integer that is greater than or equal to the ratio $(\rho
_{n}+\delta_{n})/B_{0}.$ This simple relationship makes explicit the
discontinuity that results from phase resetting in our model.

\subsection{Modes and thresholds.}

We distinguish between two primary dynamic modes for the loop-pacer system:
The \textit{reentry mode} where an action potential circulates in the loop by
itself and the \textit{paced mode} where the action potential in the loop
comes from the pacer. The system's mode in each beat changes on the basis of
threshold relations to be defined below. The modes are defined by systems of
partial difference equations that track the circulating pulse in both space
and time. Introductory material on ordinary and partial difference equations
can be found in [9, 17, 31, 40, 51]. Technically, the dynamical system
consisting of the loop, the pacer and the associated thresholds is a polymodal
structure in the sense of [51]. 

\subsubsection{\textbf{Dynamic activation duration}.}

Let $DI_{i,n}$ be the DI of Cell $i$ in beat $n.$ Define the \textit{firing
indicator} function as
\begin{align*}
\phi_{1,n} &  =1,\quad\text{and}\\
\phi_{i,n} &  =\left\{
\begin{array}
[c]{ll}%
1, & \text{if }DI_{i,n-1}>DI^{\ast}\text{ and }\phi_{i-1,n}=1\\
0, & \text{if }DI_{i,n-1}\leq DI^{\ast}\text{ or }\phi_{i-1,n}=0\text{ }%
\end{array}
\right.  ,\ i=2,\ldots,m.
\end{align*}

This quantity specifies conditions under which the $i$-th cell fires an action
potential in beat $n$; these conditions simply require that (a) the preceding
Cell $i-1$ fired and (b) the DI of Cell $i$ in the preceding beat $n-1$
exceeded the minimum value $DI^{\ast}$ as is required for propagation.

We assume that Cell 1 fires in every beat whether by a reentrant pulse or by a
pulse from the pacer.

Let $APD_{i,n}$ denote the action potential duration of Cell $i$ in beat $n$
so that if the $i$-th cell does not fire in beat $n$ then $APD_{i,n}=0$
(otherwise, $APD_{i,n}=A_{i}(DI_{i,n-1})$). The dynamic \textit{activation
duration} is defined as
\begin{equation}
AD_{i,n}=\phi_{i,n}\,A_{i}(DI_{i,n-1}),\ i=1,2,\ldots,m\nonumber
\end{equation}
where $A_{i}$ is the APD restitution function for Cell $i$. Note that
$AD_{1,n}=APD_{1,n}=A_{1}(DI_{1,n-1})$ for Cell 1 in every beat. In the
following sections (except in Case-study 4) we assume that the loop is APD
homogeneous so that $A_{i}=A$ for all $i=1,2,\ldots,m$ where $A$ is given by
(\ref{apd-shd}).

\subsubsection{\textbf{Paced mode}.}

In this mode the pacer determines the DI values $DI_{i,n}$ rather than a
reentrant pulse in the primary loop; the circulation is assumed unidirectional
owing to an active UB. We use the following system of $m$ equations:%

\begin{subequations}
\label{udc11}%
\begin{align}
DI_{1,n}  &  =\beta_{k_{n}}-\rho_{n}\label{udc1}\\
DI_{i,n}  &  =DI_{i-1,n}+AD_{i-1,n}-AD_{i,n}+CT_{i-1,n+1}-CT_{i-1,n}%
\label{udc4}\\
&  \quad+(1-\phi_{i,n})DI_{i,n-1}-(1-\phi_{i-1,n})DI_{i-1,n-1}\nonumber\\
\text{with \ }i  &  =2,\ldots,m.\nonumber
\end{align}
\end{subequations}

Here $C_{i}$ is the CT restitution function for Cell $i.$ The above system of
equations determines the DI values using the first pacer beat after time
instance $\rho_{n}$ (beat $n$ in progress). Figure 4 illustrates equations
(\ref{udc4}) and (\ref{re2}). The first five terms in (\ref{udc4}) simply
express in mathematical terms what we might state less precisely using the
English language. They are the same for tissue fibers as they are for loops;
see, e.g. [68]. The last two terms in the definition of $DI_{i,n}$ in
(\ref{udc4}) allow for proper adjustment of the DI value (without double
counting) when Cell $i$ does not fire in beat $n$ (see comments on the reentry
mode below). In typical loop models in prior literature all cells fire so
these two terms would drop out.

\begin{figure}[ptb]
\begin{center}
\includegraphics[width=3.5in]{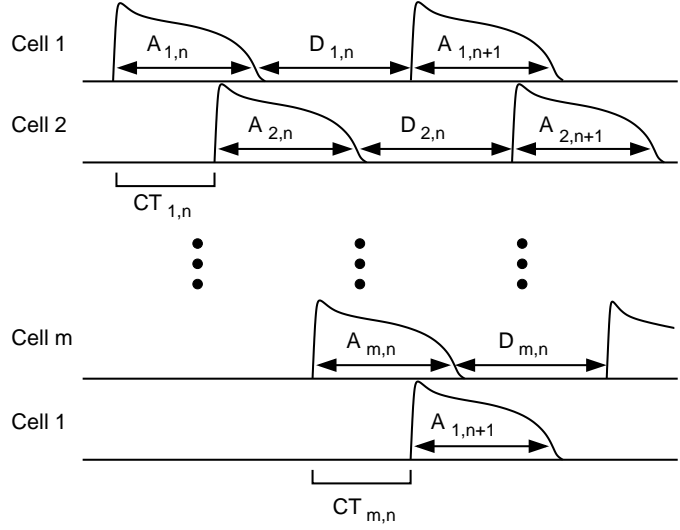}
\end{center}
\caption{Schematic diagram of a circulating action potential in a ring of $m$
cells; A is APD, D is DI and CT is the conduction time. The sizes of A, DI and CT are selected arbitrarily for clarity of illustration.}%
\end{figure}

\subsubsection{\textbf{Reentry mode}.}

In this mode the reentrant loop drives the ventricles ectopically. Thus
$DI_{1,n}$ is the difference between the APD of Cell $1$ and the total
conduction time for unidirectional circulation once around the ring, i.e.
\begin{subequations}
\label{re0}%
\begin{align}
DI_{1,n}  &  =\sum_{j=1}^{m}CT_{j,n}-A_{1}(DI_{1,n-1})\label{re1}\\
DI_{i,n}  &  =DI_{i-1,n}+AD_{i-1,n}-AD_{i,n}+CT_{i-1,n+1}-CT_{i-1,n}%
\label{re2}\\
&  \quad+(1-\phi_{i,n})DI_{i,n-1}-(1-\phi_{i-1,n})DI_{i-1,n-1}\nonumber\\
\text{with \ }i  &  =2,\ldots,m.\nonumber
\end{align}
\end{subequations}

It is not hard to show that the system of $m$ equations (\ref{re0}) is
equivalent to the coupled map lattice equations in [51] when $\phi_{i,n}=1$
for all $i$ and $n.$

Equations (\ref{re0}) are not valid if some of the DI\ values fall below the
threshold $DI^{\ast}$ or if a stray pacer pulse gets into the loop (i.e. if
the right hand side of (\ref{re1}) exceeds the right hand side of
(\ref{udc1})). The following threshold relations ensure a consistent scenario.

\medskip

\begin{quote}
TC: The \textit{circulation threshold }(or \textit{head-tail }or\textit{
conduction block} \textit{threshold})\textit{ }is based on the following set
of inequalities:
\begin{align*}
\sum_{j=1}^{m}CT_{j,n}-A_{1}(DI_{1,n-1}) &  >DI^{\ast},\\
DI_{i-1,n}+AD_{i-1,n}-AD_{i,n}+CT_{i-1,n+1}-CT_{i-1,n} &  >DI^{\ast}%
\quad\text{for}\ i=2,\ldots,m.
\end{align*}

They ensure that the DI values $DI_{i,n}$ in Equations (\ref{re0}) exceed $DI^{\ast},$
as is necessary for the generation of action potentials. If any of the above
inequalities fails in a Cell $i$ then a wavefront reaches its own tail in that
particular cell and circulation stops until the arrival of the next pacer
pulse. In this case we say that TC fails in Cell $i.$
\end{quote}

\medskip

\begin{quote}
TP: The \textit{pacer} \textit{threshold} (or \textit{phase resetting}
\textit{threshold}) is defined by the following inequality:%
\[
\sum_{j=1}^{m}CT_{j,n}-A_{1}(DI_{1,n-1})<\beta_{k_{n}}-\rho_{n}.
\]

This inequality ensures that the reentrant pulse (rather than a stray one from
the pacer) reactivates Cell 1.
\end{quote}

\medskip

\textbf{Mode switching criteria}. \textit{Equations (\ref{re0}) are used if TP
holds and also TC holds in Cell 1. If for some Cell }$i\geq2$\textit{ TC
fails, then }(\ref{re2}) \textit{can still be used to find the DI values for
beat n, but we switch to equations (\ref{udc11}) for beat }$n+1$\textit{. If
TC fails for }$i=1,$ \textit{or if TP fails then we use (\ref{udc11}) for beat
n.}

\subsubsection{\textbf{Bidirectional circulation (BDC) mode}.}

In the absence of unidirectional block the pacer sends two pulses in the loop
with Cell 1 activating both Cell 2 and Cell $m$. These pulses propagate in
opposite directions to eventually annihilate each other in some cell in the
loop. Transitions into and out of this bidirectional circulation mode are
governed by TC, TP and a UB threshold (TUB); see Section 4 below. We assume 
for simplicity that TUB never fails so that the BDC mode need
not be discussed in this paper.

\section{SITR patterns.}

In this section we present the results of numerical simulations of our
threshold model along with some analytical observations. Due to the large
number of different types of behavior that can occur by changing one or more
of the many variables involved, providing a comprehensive list of essentially
different patterns is not feasible here. We therefore present our main results
in the form of a few model case-studies and for additional results we refer
the interested reader to our web site: \textit{www.people.vcu.edu/\symbol{126}%
hsedagha/SITR}.

Each case-study below consists of one or more \textquotedblleft
runs,\textquotedblright\ i.e. sets of iterations of the propagation equations.
Each run is based on a fixed set of parameter values and we call each
iteration in a given run a \textquotedblleft beat\textquotedblright. For each
run we need to specify $m$ initial DI values $DI_{i,0}$. This specification of
DI\ values is a technical representation of the occurrence of a premature
stimulation in our model.

Computational accuracy is not a major issue in the qualitative studies in this
paper. However, calculating the values of the various nonlinear functions (e.g. the
exponentials in the APD and CV\ restitution functions) generate round-off errors
that can accumulate in certain cases (e.g. runs in Case-study 4). Therefore, results
obtained by different programs, codes or levels of precision (single or
double) may be different from some of those that we have presented, although the
qualitative features are usually retained.
\subsection{\textbf{Mode sequences}.}

The following conventions facilitate the presentation of results in this section:

\begin{quote}
For a loop consisting of $m$ cells if a reentrant pulse is blocked at Cell $j$
in a particular beat, then we define the \textit{reentry value} of that beat
as the ratio $(j-1)/m;$ this ratio will be abbreviated as $[\,j].$

If a reentrant pulse completes a turn around the loop then the reentry value
is 1.

If a reentrant pulse is blocked in Cell $j$ after $t$ complete turns around
the loop then the reentry value is $t[\,j].$

If a beat occurs in the paced mode (i.e. one of the thresholds TC or TP fails)
then we assign it the value $-1.$ Thus $t$ consecutive paced beats appear as $-t$.

If the pattern is locked in the reentry mode, we use the symbol $\infty;$ if
it is locked in the paced mode, we use $-\infty.$
\end{quote}

Adding up consecutive reentry mode beats and consecutive paced mode beats
gives a sequence of numbers consisting of integers whose sign tell us the mode
of system in various sets of beats. Examples of mode sequences and more
details on them are found in the case-studies below.

\subsection{Bistability and thresholds.}

We call the loop-pacer system \textit{bistable} if there are two or more
coexisting stable DI configurations or state vectors, each of which can be
reached from a particular region of the $m$-dimensional state space. If there
are more than two distinct, coexisting stable states then the system is
\textit{multistable}.

Bistability may be attributed to the non-concavity of the APD restitution
function [50]. The APD function (\ref{apd-shd}) whose graph has bumps and
twists is also non-concave. Proper CT restitution parameters are required to
realize bistability; i.e. the CT and APD restitution parameters must be
properly matched. Because the length $L$ of the loop affects the conduction
time through it, bistability may emerge or fade as $L$ is changed with other
CT parameters fixed [50].

A bistable regime can affect SITR\ patterns by causing the violation of the
circulation threshold TC without any changes in the APD or CT parameters, or
any changes in the pacing rate. This situation occurs if the bistable regime
satisfies the following \textit{threshold bistability} condition:

\medskip

\begin{quote}
\textit{T-bistability}: There are two distinct stable states in the reentry
mode: In one state the DI values cross $DI^{\ast}$ and cause the failure of
the circulation threshold TC at some cell within the loop, but in the other
state the DI values are always greater than $DI^{\ast}$.
\end{quote}

\medskip

An example of T-bistability is shown in Figure 7 below; other examples of
T-bistability occur in the next section (Case-study 1). The $m$-dimensional
state space of a T-bistable system is partitioned into two regions or basins
of attraction. As a SITR pattern evolves with the motion of the DI state
vector, mode changes occur if the state vector crosses over into the basin of
attraction of a different stable regime.

\subsection{Case-study 1: T-bistable reentry.}

In this section we establish that T-bistability lends a measure of
unpredictability to SITR\ patterns that is not due to phase resetting
disruptions by the pacer (i.e. TP\ failures). For instance, comparing the mode
sequences in Runs 1.2, 1.4 and 1.6 below we see that the initial few bursts of
fast, reentrant beats provide no obvious clues about the eventual modes that
the patterns lock into.

Consider a homogeneous loop with the APD\ and CT restitutions given by
(\ref{apd-shd}) and (\ref{CT}). We also set the length $L$ and other
parameters as:

\begin{center}%
\begin{tabular}
[c]{|c|c|c|c|c|}\hline
$L$ & $m$ & $DI^{\ast}$ & $\delta$ & $B_{0}$\\\hline\hline
12.5 cm & 125 & 15.3 ms & 120 ms & 320 ms\\\hline
\end{tabular}

\end{center}

\noindent where $B_{0}$ is the fixed cycle length of the pacer; it is chosen
small in this case-study to facilitate re-initiations of reentry. The value of
$\delta$ is set high enough to inhibit pacer interference (TP never fails).
The following table summarizes the results of simulations; it is followed by a
series of observations and elaborations.

\begin{center}%
\begin{tabular}
[c]{|c|c|c|}\hline
Run No. & $DI_{i,0},\ i=1,\ldots125$ & SITR pattern mode
sequence\\\hline\hline
1.1 & 200 ms & $\{-2,2,-2,2,\ldots\}$\\\hline
1.2 & 100 ms & $\left\{  -2,4,-2,32[73],-2,\infty\right\}  $\\\hline
1.3 & 80 ms & $\{-1,\infty\}$\\\hline
1.4 & 70 ms & $\left\{  4,-2,14,-2,4,-2,22[49],-\infty\right\}  $\\\hline
1.5 & 60 ms & $\{\infty\}$\\\hline
1.6 & 50 ms & $\left\{  11,-2,4,-2,27[57],-2,\infty\right\}  $\\\hline
\end{tabular}

\end{center}

(i) In Run 1.1 the SITR pattern immediately locks into a regular form of 2
paced beats followed by 2 reentrant beats. In Runs 1.2, 1.3, 1.5 and 1.6 the
SITR\ pattern locks into the reentry mode.\ In Run 1.4 the paced mode is
locked into. Note that the only thing that is changing from one run to the
next is the initial DI values.

(ii) In Runs 1.2, 1.3, 1.5 and 1.6 the DI values in the sustained or locked
reentry modes converge to a fixed number and the eventual values of the
fundamental parameters are

\begin{center}%
\begin{tabular}
[c]{|c|c|c|}\hline
DI & APD & cycle length CL = APD+DI\\\hline\hline
59.4 ms & 173.6 ms & 233 ms\\\hline
\end{tabular}

\end{center}

The fixed, limiting DI value 59.4 represents a stable \textit{convergent
state}. This equilibrium DI is locally stable (attracting) because using the
restitution functions (\ref{apd-shd}) and (\ref{CT}) we compute [4, 26]%
\[
A^{\prime}(59.4)+C^{\prime}(59.4)=0.90<1.
\]

(iii) In Runs 1.2, 1.4 and 1.6 all the spontaneously terminated reentry bursts
have oscillating (quasiperiodic) DI. Thus APD\ and CL also oscillate similarly. The
oscillatory state is stable and eventually crosses $DI^{\ast}=15.3$ ms. Hence
there is T-bistability (the oscillatory one and the convergent one) which is 
responsible for mode changes in each run.

For example, in Run 1.2 the 2nd initiation of reentry (the 32-beat burst)
starts with the DI\ state-vector in the basin of attraction of an oscillatory
state which expands enough for the DI\ values to cross $DI^{\ast}$ (TC fails);
see Figure 5. This terminates the second reentry burst at Cell 73 in the 41st
beat. This is the same type of termination mechanism as that in [19, 20, 26];
this type of oscillations in DI are also seen in [56] where they are related
to the T wave alternans in ECG recordings; also see [35] in this regard. With
the 3rd initiation of reentry, the DI state-vector falls in the basin of
attraction of the convergent state which then effectively \textquotedblleft
traps\textquotedblright\ the DI trajectory and locks the SITR pattern into the
reentry mode (here then, simplicity of behavior is not a good thing).

\begin{figure}[ptb]
\label{run1-2}
\par
\begin{center}
\includegraphics[width=4.5in]{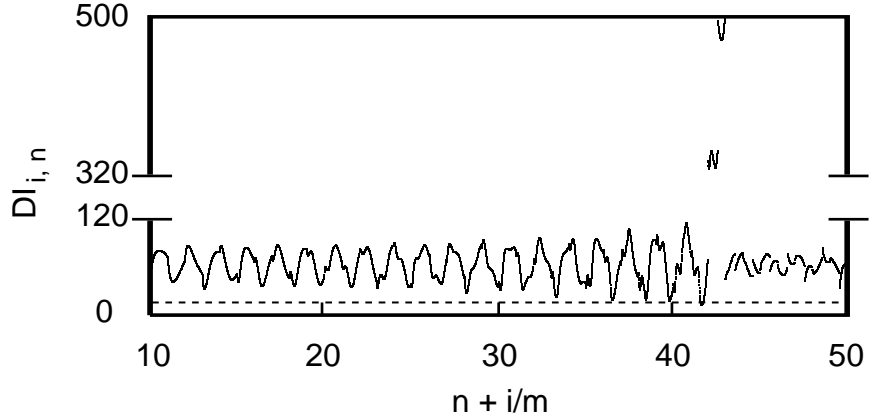}
\end{center}
\caption{\small Results of Run 1.2, showing variation of DI in Beats 10 through 50.
The dashed line corresponds to $DI^{\ast}=15.3$. Amplification of oscillations
in DI leads to termination of reentry in Beat 41, Cell 73 because
$DI_{73,41}<DI^{\ast}$. With Beat 44 reentry is re-initiated and the system locks 
into the stable convergent state (after some transient oscillations).}%
\end{figure}

(iv) Examination of the simulation data for Run 1.4 shows a recurrent failing
of TC in Cell 1 after the last burst of 23 reentrant beats. For all
sufficiently large values of $n$,

\begin{center}%
\begin{tabular}
[c]{|c|c|c|}\hline
$DI_{1,n}$ & $APD_{1,n}$ & $CL_{n}$\\\hline\hline
96.6 & 223.4 & 320\\\hline
\end{tabular}

\end{center}

The number 96.6 is the equilibrium DI for the paced mode with pacing period
$B_{0}=320$. The paced-mode equilibrium happens to be
stable because in the absence of reentry the paced mode is
governed by the one-dimensional difference equation%
\[
DI_{1,n}=F(DI_{1,n-1})=B_{0}-A(DI_{1,n-1})
\]
with $F^{\prime}(96.6)=-A^{\prime}(96.6)=-0.901;$ i.e. $|F^{\prime}(96.6)|<1.$
Why the transition to the stable paced equilibrium occurs in this run and not
others is unclear. Figure 6 shows the changes in DI values in this run.

\begin{figure}[ptb]
\label{run1-4}
\par
\begin{center}
\includegraphics[width=4.5in]{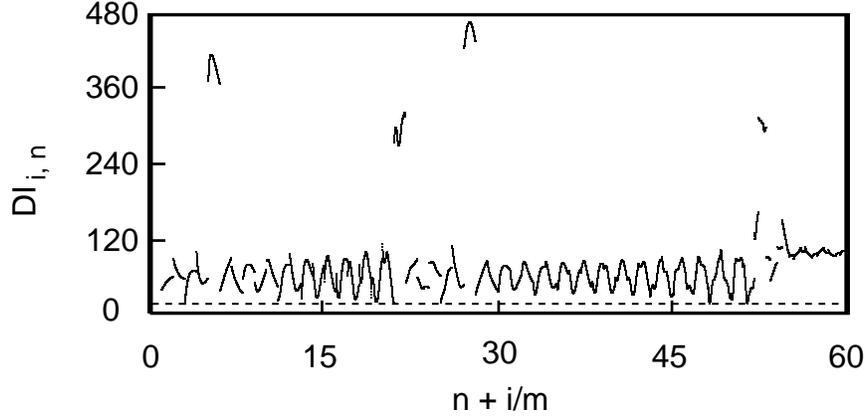}
\end{center}
\caption{\small Results of Run 1.4, showing variation of DI in Beats 1 through 60.
The dashed line corresponds to $DI^{*} = 15.3$. The four interruptions of reentry
due to conduction block (TC fails) are seen as brief jumps to high DI. The last 
interruption terminates reentry and locks the system in the paced mode.}%
\end{figure}

\medskip

(iv) In Run 1.5, the given initial DI values put the state vector in the basin
of attraction of the convergent state of the system. This is evidently not the
case in Run 1.6 with even shorter DI. In Run 1.6, as in Run 1.4, the initial
mode is oscillating-DI reentry.

\subsection{Case-study 2: Non-T-bistable reentry.}

We now consider some longer loops than in Case-study 1 for comparison. In
these loops bistability is present but does not satisfy the T-bistability
condition. A smaller variety of different SITR\ patterns are obtained (with
all other system parameters having the same values as in Case-study 1). We
consider two different lengths:

\begin{quote}
$L=13$ cm ($m=130$) where stable oscillating states exist all of which fail
TC, but the convergent mode is absent since the DI fixed point of about $62.5$
ms is unstable [4, 26]%
\[
A^{\prime}(62.5)+C^{\prime}(62.5)=1.013>1.
\]

$L=14$ cm ($m=140$) where all of the stable oscillating states (there is at
least one) satisfy TC. The convergent mode is again absent for the same
reason as above:%
\[
A^{\prime}(68)+C^{\prime}(68)=1.253>1.
\]

\end{quote}

To save space let us abbreviate repetitions or locked mode patterns with bars:%
\begin{equation}
\overline{k,-j}=k,-j,k,-j,\ldots\label{brac-bar}%
\end{equation}

The following table summarizes the results of simulations; it is followed by a
series of observations and elaborations.

\begin{center}%
\begin{tabular}
[c]{|c|c|c|c|}\hline
Run No. & $DI_{i,0}$ & $L=13$ cm & $L=14$ cm\\\hline\hline
2.1 & 200 ms & $\left\{  \overline{-2,2}\right\}  $ & $\{-2,2,-2,\infty
\}$\\\hline
2.2 & 100 ms & $\left\{  -2,4,-2,12[60],\overline{-2,2}\right\}  $ &
$\{-2,\infty\}$\\\hline
2.3 & 80 ms & $\left\{  \overline{2,-2}\right\}  $ & $\{-1,1,-2,2,-2,\infty
\}$\\\hline
2.4 & 70 ms & $\left\{  12,-2,10[79],-2,17[79],\overline{-2,5[79]}\right\}  $
& $\{12,-2,2,-2,\infty\}$\\\hline
2.5 & 60 ms & $\left\{  56,\overline{-2,2}\right\}  $ &
$\{-1,10,-2,9,-2,\infty\}$\\\hline
2.6 & 50 ms & $\left\{  3,-2,4,-2,12[60],\overline{-2,2}\right\}  $ &
$\{-\infty\}$\\\hline
\end{tabular}

\end{center}

(i) The SITR patterns for the 14 cm loop are expected to lock into the reentry
mode since the stable states do not fail TC to cause terminations. The
exception is Run 2.6 where possibly owing to the greater length of the loop, a
pulse from the pacer has time to activate Cell 1 and block the premature
simulation from reentering the loop (threshold TP fails). It is not clear why
reentry does not re-initiate as in Run 2.5.

(ii) In the case of the 13 cm loop the repeated terminations can be attributed
to the fact that the stable states all fail TC; thus the patterns cannot lock
into the reentry mode.

(iii) For both of these loops, the eventual form of the SITR\ pattern is more
predictable than the T-bistable loop of Case-study 1. On the other hand, the
transient bursts of fast beats still seem difficult to predict.

\subsection{Case-study 3: Special patterns.}

\textbf{Run 3.1} (\textit{more T-bistability}). This run presents a complex
SITR pattern that locks into a long cycle of several initiations and
terminations. Make the following changes to parameters in Run 1.2:

\begin{center}%
\begin{tabular}
[c]{|c|c|}\hline
$d$ & $B_{0}$\\\hline\hline
1.3 & 315 ms\\\hline
\end{tabular}

\end{center}

Note that the increase in $d$ from 1 to 1.3 in this run causes a slight
elevation of the CT restitution curve (i.e., a conduction slow-down) at low DI
values; see Figure 3.

(i) In this run, after a transient period of spontaneous initiations and
terminations, a long 62-beat pattern emerges that is repeated; i.e. the
SITR\ pattern locks into a 62-beat cycle containing several bursts of fast
reentrant beats. The mode sequence for this run is as follows, with the
pattern between starred numbers repeating:
\begin{align*}
&  \{-2,8[81],-1,[35],-2,4,-2,7[24],-1,[18],-2,8[18],^{\ast}%
-1,[12],-2,7[12],-1,[12],-2,\\
&  10[49],-2,5[49],-2,4,-2,7[22],-1,[19],-2,7[19],^{\ast}%
-1,[12],-2,7[12],-1,[12],-2,\ldots\}
\end{align*}

(ii) The DI equilibrium of approximately $64$ ms is unstable in this run since%
\[
A^{\prime}(64)+C^{\prime}(64)=1.073>1.
\]

This instability is caused by the greater value of $d$ and results in the
absence of the convergent state. Nevertheless, there is T-bistability where
one of the stable oscillating states does not fail TC but another oscillating
state does. In the first oscillatory state, DI values come very close to
$DI^{\ast}$; see Figure 7. This proximity increases the sensitivity to
transient effects of mode changes and makes locking into the reentry mode an
improbable event.

\medskip

\begin{figure}[ptb]
\label{run3-1}
\par
\begin{center}
\includegraphics[width=5.5in]{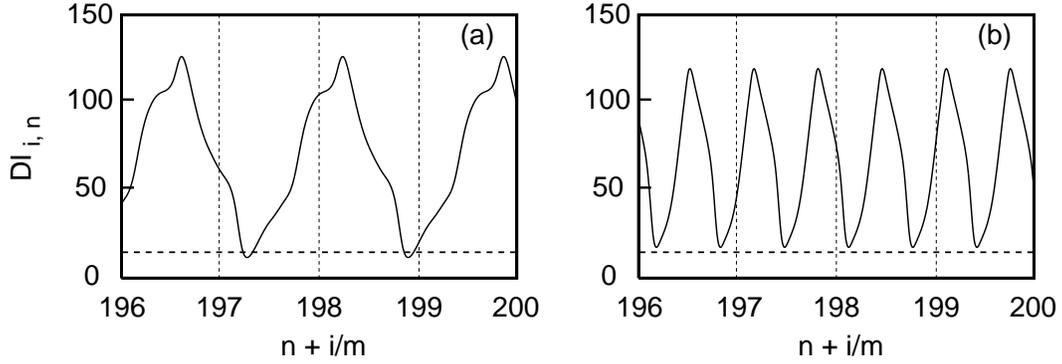}
\end{center}
\caption{\small Two distinct stable states illustrating T-bistability in Run 3.1. 
(a) If $DI_{i,0}= 67$ for all $i$, the graph crosses the dashed horizontal line $DI = 15.3$;
for the sake of illustrating the quasiperiodic nature of oscillations in reentry mode, the graph ignores TC since reentry would terminate with the first such crossing. (b) $DI_{i,0} = 66$ for all $i$. Reentry is sustained with quasi-periodic variation of DI.}%
\end{figure}

\medskip

\textbf{Run 3.2} (\textit{pacer action}). In Case-studies 1 and 2 we see that the
convergent reentry pattern is \textquotedblleft sticky\textquotedblright\ i.e.
once this pattern is attained, reentry does not self-terminate by the
circulation mechanism in the loop. However, pacer interaction with the loop
can lead to termination in some cases by changing the convergent pattern to a
terminating oscillatory one in a T-bistable case.

All parameters are as in Run 1.5 (hence there is T-bistability) except that now:

\begin{center}%
\begin{tabular}
[c]{|c|c|}\hline
$\delta$ & $B_{0}$\\\hline\hline
45 ms & 800 ms\\\hline
\end{tabular}

\end{center}

We have set the pacing period at the nominal sinus length to eliminate the
effects of fast pacing. The smaller value of $\delta$ promotes interference by
the pacer.

(i) The mode sequence for this run, which may be compared with Run 1.5, is:%
\[
\{40,-1,15,-\infty\}\text{.}%
\]

(ii) Reentry is initiated with Beat 1 by a premature stimulation, and the DI
values begin to approach the limiting value 59.4 ms as in Run 1.5. However, TP
fails in Beat 41; a stray pulse from the pacer enters the loop and changes the
reentry DI pattern from the convergent type to an oscillating one; see Figure 8.
The quantities $PL_{n}$, $PDI_{1,n}$ in Figure 8 are the right hand sides of 
equations (\ref{udc1}) and (\ref{re1}), respectively. 

\medskip

\begin{figure}[ptb]
\label{run4-1}
\par
\begin{center}
\includegraphics[width=5.5in]{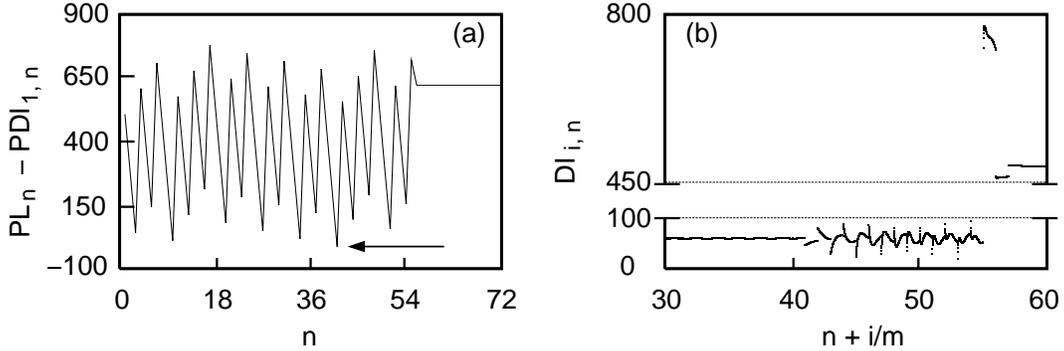}
\end{center}
\caption{\small Results of Run 3.2. (a) Variations in $PL_{n} - PDI_{1,n}$ (see text). 
The negative value in Beat 41 (indicated by the arrow) corresponds to
termination of reentry due to failure of TP. (b) DI values in Beats 30 through 60; note the phase
resetting that occurs in Beat 41 and changes the reentrant circulation pattern from convergent 
to oscillatory in this T-bistable case.}%
\end{figure}

The oscillating DI pattern crosses the $DI^{\ast}$ threshold (TC fails) in
Beat 55 and reentry terminates in this beat.

\subsection{Case-study 4: Slow pacing, heterogeneity and PVC's.}

When the pacing period is long (say, within the nominal sinus range of
800$\pm350$ ms) and the loop is homogeneous, \textit{re-initiations} of
reentry do not occur with the above CT and APD parameters in the absence of
premature stimulations or ectopic sources. The low CT (or high CV) in
the loop blocks reentry when pacing is slow. Nevertheless, the rather common occurrences of
ventricular tachycardia at low pacing rates in patients [58] motivate us to
study this important special case in the context of a heterogeneous loop with
a variable-rate pacer.

Suppose that the loop consists of two patches of cells. Patch 1 consists of
cells 1 through $j$ and Patch 2 contains cells $j+1$ and beyond where of
course, $j$ is a positive integer less than $m,$ the total number of cells.
The cells in Patch 1 may have different APD and/or CT restitution parameters
than those in Patch 2. For simulations, we also assume that the pacer has
a variable-rate within the aforementioned nominal sinus range. Assume that the pacer's
oscillation period varies in a sinusoidal fashion:%
\[
B(k)=B_{0}+B_{1}\cos\left(  \frac{2\pi k}{w}\right)  .
\]

This beat pattern may represent an idealization (without random effects) of a
single cycle of a complex, multi-pattern stretch of pacer beats. More complex
variations may be considered in later studies if needed.

The number $w$ is the period of a full cycle of variable-rate oscillations,
which we refer to as a \textit{full pacing cycle} (FPC). The halfway point of
the FPC, or the \textquotedblleft bottom of the FPC well\textquotedblright%
\ occurs at $k=w/2;$ at this point, the pacer oscillates with minimum period
(fastest beat rate). On the other hand, when $k=0$ or $k=w$ the pacer has its
largest oscillation period (slowest beat rate). The number $B_{1}$ is the FPC
\textquotedblleft amplitude\textquotedblright\ since it modulates the pacer's
oscillation period within the FPC. If $B_{1}=0$ then the period is a fixed
$B_{0}$ as in the preceding case-studies.

\medskip

\textbf{Run 4.1}. We fix the APD and CT parameters for Patch 2 to be the same
as those used in Case-study 1. Patch 1 APD and CT have the same parameters as
Patch 2 except for the following:

\begin{center}%
\begin{tabular}
[c]{|c|c|c|}\hline
$a_{1}$ & $a_{2}$ & $c$\\\hline\hline
300 ms & 170 ms & 0.05 cm/ms\\\hline
\end{tabular}

\end{center}

Thus APD is reduced (by a shift in the restitution function) in Patch 1; also
the cells in this patch conduct more slowly with a lower value of $c$ (the maximum CV).
Additional loop and pacer parameters are set as follows:

\begin{center}%
\begin{tabular}
[c]{|c|c|c|c|c|c|c|}\hline
$L$ & $B_{0}$ & $B_{1}$ & $w$ & $j$ & $\delta$ & $DI_{i,0},\ 1\leq i\leq
135$\\\hline\hline
13.5 cm & 800 ms & 350 ms & 1000 & 50 & 140 ms & 600 ms\\\hline
\end{tabular}

\end{center}

The integer $j$ gives the number of cells in Patch 1 and the three pacing
parameters $w,B_{0}$ and $B_{1}$ define a variable pacing protocol with FPC
period of 1000 beats and amplitude 350 about a fixed, nominal sinus trend of
800 ms.

(i) The mode sequence of the SITR pattern obtained is%
\[
\{-464,[51],-1,[51],-1,[51],-1,[51],-1,[51],-2,\infty\}.
\]

(ii) The mode sequence exhibits a pattern of 5 incomplete reentry initiations
followed by a pacer beat (or 2 in one case). Each incomplete reentry start
which is blocked at the patch junction (Cell 51) corresponds to a short cycle
length (about 224 ms) compared to the pacer's period of approximately 460 ms
near the middle of the FPC. Hence the pattern of 5 alternating reentry/pacer
beats resembles a series of premature ventricular contractions (PVC) in a
bigeminal form before reentry finally takes hold for good.

(iii) We note that if $j=40$ or fewer, then no reentry is initiated in this run. 
On the other hand, larger $j$ than 50
sustain reentry more quickly by increasing the percentage of slow cells in the loop.

\medskip

\textbf{Run 4.2}. In the preceding run the pacer did not interfere with the
reentrant circulation (TP did not fail) because $\delta=140$ ms was large
enough to block the pacer. We now show that decreasing $\delta$ to promote
pacer interactions may cause the eventual termination of reentry in the
preceding run. We make only one change in this run relative to Run 4.1 by
setting $\delta=70$ ms.

(i) The mode sequence of the resulting SITR pattern is as follows:%
\begin{align*}
&  \{-464,\underset{\text{4 times}}{\underbrace{[51],-1,\ldots,[51],-1}%
},[51],-2,1,-1,11,-3,1,-1,4,-1,6,-3,1,-1,\\
&  4,-1,6,-3,1,-1,11,-1,6,-3,\underset{\text{38 times}}{\underbrace
{[51],-1,\ldots,[51],-1}},[51],-374,\ldots\}
\end{align*}

After several PVC's and reentrant bursts of different durations, the pattern
locks into the paced mode until reentry is initiated in the next FPC. This
mode sequence is sensitive to computer round-off errors; however, the varied
nature of mode changes persists. Figure 9 gives a graphical representation of
mode changes for this run.

\begin{figure}[ptb]
\begin{center}
\includegraphics[width=5.5in]{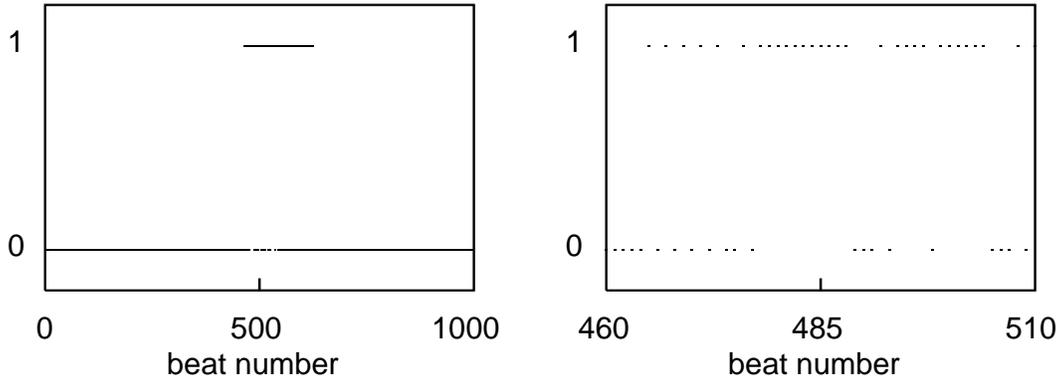}
\end{center}
\caption{\small Schematic diagram of mode changes in Run 4.2. Beats in the reentry mode 
are listed with value 1 and those in the paced mode are given the value 0.}%
\end{figure}

(ii) The first deviation from the mode sequence of Run 4.1 occurs with a TP
failure that interrupts reentry after just one full cycle. Some of the $-1$
entries in the mode sequence above (mostly the ones \textit{not} associated
with conduction blocks in Cell 51) indicate failures of TP also.

\section{Further details of the model.}

In this section we present additional supporting details about the model.

\subsection{Remarks concerning the APD\ restitution function.}

\subsubsection{Memory and latency. }

Experimental measurements and other studies reveal the difficulty of fitting a
single curve to the data, particularly at the lower end of the DI scale. To
address this and related issues, refinements have been considered in the forms
of memory and latency. Beat to beat memory has been studied in [13, 19, 21,
23, 28, 46, 56] and has been modeled in a number of ways. In many cases,
modeling memory can be accomplished by using DI\ values from several earlier
stimulations. Latency (roughly a brief period between the arrival of a pulse
and the firing of the cell) has also been expressed as a restitution function
of DI [12, 13]. If this effect is added to APD we obtain a composite
restitution function that may no longer be monotonic for small DI values. For
simplicity, we ignore both memory and latency in this paper; however, they can
(and need to) be included in the model for greater refinement in future studies.

\subsubsection{Standard vs. dynamic protocols.}

The APD restitution function may be measured experimentally using two
different pacing protocols: Dynamic or standard (S1-S2) [29, 32]. These
protocols generally produce APD curves that are shifted or have different
slopes. For a discussion of this issue regarding the APD restitution that we
use in this paper see [33]. The fit in (\ref{apd-shd}) is to the dynamic
restitution data in [33].

\subsubsection{Remodeling.}

The APD values (and thus the restitution function) may be affected by the
steady fast pacing of the ventricles [34] or of the atria [39]. The implied
shift in the APD restitution can have a significant effect on the
SITR\ patterns in certain borderline cases. For simplicity we do not consider
such effects in this paper; however, a suitable modification of the 
APD restitution function that
incorporates explicit time dependence can accommodate dynamically-induced shifts.

\subsubsection{APD non-monotonicity and spatio-temporal chaos.}

Experimental evidence is offered in [53] that the APD restitution curve may be
non-monotonic for small DI values. Working with a APD restitution curve having
a single local minimum at low DI, the same study demonstrates the occurrence
of spatial bifurcations along the length of a periodically paced fiber of
cardiac tissue. Although not involving a loop structure, the work in [53] uses
restitution functions in a discrete model. A localized form of
non-monotonicity (like a bump or dent) in APD may occur at any DI value and
has interesting consequences for additional complexity in reentry mode. In
[48] where a continuous model of the loop as a cable is used, this type of
non-monotonicity is shown to be responsible for the appearance of
spatio-temporal chaos in APD and other key variables over the length of the
loop. The occurrence of spatio-temporal chaos on the loop may raise the level
of unpredictability in SITR if the amplitudes of oscillations in DI are large
enough to let the wavefront reach its tail (thus causing a threshold failure).

\subsection{\textbf{Restitutions of ionic currents and conduction block}.}

Various anatomical and functional mechanisms for UB in a loop are discussed in
the model study [49]. These mechanisms include a properly
timed premature stimulus delivered to a suitable location in the loop,
inhomogeneities in the degree of cellular uncoupling and gap junction
resistance, in membrane excitability and in fiber cross sectional area. The
discussion in [49] points to time, space and voltage
\textquotedblleft windows of vulnerability\textquotedblright\ for the
occurrence of unidirectional block and initiation of reentry. Of these, the
time window is seen to be the easiest to measure and its value is related to
the space and voltage windows via standard mathematical formulas.

Each cell requires a certain amount of depolarizing current to fire an action
potential. This \textit{activation threshold} can be represented by a
restitution function [27] or equivalently, the strength-interval curve [13]
shifted to the left so that the origin represents the end of the ERP.  In
Figure 10 this curve is shown as a decreasing restitution function $I_{a}$.

In principle, at any time following the end of the ERP a sufficiently large
current will be able to elicit an action potential. However, in the absence of
external electrical shocks, there is a limited amount of current $I_{g}$
available to depolarize the next cell. If a restitution of $I_{g}$ is given
then the intersection point of $I_{a}$ and $I_{g}$ is the $DI$ threshold or cut-off
value $DI^{\ast}$ in the sense that if $DI>DI^{\ast}$ then action potential is
fired and propagation occurs.
The time interval $DI^{\ast}$ is similar in nature to the chronaxie\ [47] although
the definition of $DI^{\ast}$ requires another restitution curve in addition
to $I_{a}.$ A monotonically increasing restitution function for $I_{g}$ might
be appropriate; see Figure 10.

\begin{figure}[ptb]
\label{currents}
\par
\begin{center}
\includegraphics[width=3.0in]{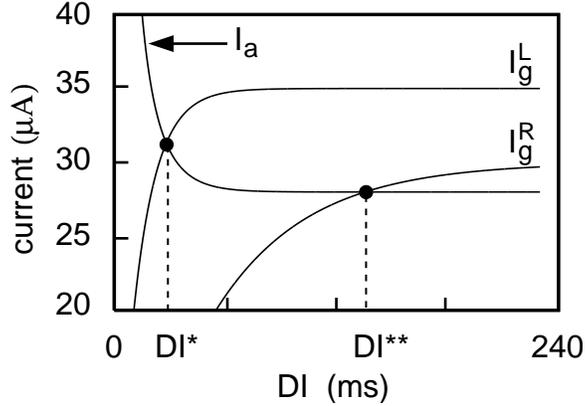}
\end{center}
\caption{\small Activation and generated currents restitution curves
illustrating the definition of $DI^{\ast}$ and $DI^{\ast\ast}$. In this
figure, $I_{a}=I_{a}^{L}=I_{a}^{R}$ (see text).}%
\end{figure}

For our purposes here we do not need explicit formulas for either $I_{a}$ or
$I_{g}.$ The value $DI^{\ast}$ may be chosen
arbitrarily or estimated experimentally without reference to its potential
sources. Values as high as 160 ms were used in
[26]. In human cases the value of $DI^{\ast}$ in vivo is likely to be in a
much smaller range. We use the conservatively small value of about 15 ms in
this study.

An analog of the time window in [49] can be defined here using the restitution
functions $I_{a}$ and $I_{g}$. We define two sets of currents restitution
functions for Cell 1 corresponding to the two possible directions in the loop:
$I_{a}^{L}$, $I_{a}^{R}$ and $I_{g}^{L}$, $I_{g}^{R}$ for the
\textquotedblleft left\textquotedblright\ and the \textquotedblleft
right\textquotedblright\ directions. The asymmetry creates a \textit{UB
window} in Cell 1 as follows: Let $I_{a}^{R}\geq I_{a}^{L}$ and $I_{g}^{R}\leq
I_{g}^{L}$ with at least one of these inequalities strict. Let $DI^{\ast}$ be
the (unique) intersection point of $I_{a}^{L}$ and $I_{g}^{L}$ and let
$DI^{\ast\ast}$ to be the intersection point of $I_{a}^{R}$ and $I_{g}^{R}.$
Then $DI^{\ast}<DI^{\ast\ast}$ and the interval between these two values is
where the UB occurs. For if $DI_{1,n}$ represents the DI of Cell 1 in a cycle
or beat $n,$ then the conditions
\[
DI^{\ast}<DI_{1,n}<DI^{\ast\ast}%
\]
imply that propagation is possible in one direction because the threshold
marked by $DI^{\ast}$ is crossed but it is inhibited in the other direction
where the $DI^{\ast\ast}$ threshold is not crossed.

We emphasize that the unidirectional block as defined here is functional and
may not occur even if the UB window exists (for anatomical or functional
reasons). If the number $DI_{1,n}$ does not enter the the UB window (i.e. the
interval from $DI^{\ast}$ to $DI^{\ast\ast}$) in a particular beat $n$ then
the UB\ does not materialize. This option adds another mode to the loop-pacer
system, namely the BDC mode mentioned previously. In addition to TC and TP,
the following threshold is associated wtih the BDC mode:

\medskip

\begin{quote}
TUB: $DI_{1,n-1}<DI^{\ast\ast}$ with beat $n-1$ in the paced\ mode.
\end{quote}

This inequality defines the \textit{UB threshold} to ensure that Cell 1
conducts unidirectionally. If beat $n-1$ is in reentry mode then
conduction in beat $n$ is automatically unidirectional whether TUB holds or
not, because the pulse is transmitted from a still active Cell $m.$ To limit
the variety of cases to consider, we assume in this paper that $DI^{\ast\ast}$
is so large that the unidirectional block is always present in Cell 1
(permanent UB) so TUB always holds.

\section{Summary and discussion.}

In the preceding sections, we developed an interactive loop-pacer threshold
model that is a part of a familiar cardiac anomaly. Using experimental data
on restitution parameters from the existing literature for action
potential duration and conduction time, we obtained various conditions and
parameter values that would cause irregular patterns of spontaneous
initiations and terminations of reentry.

A major cause of unpredictable behavior is threshold bistability where one 
stable regime causes a failure of the circulation
threshold TC and another does not. In this situation the initiation
and termination of reentry depends on the location of the DI state vector
in the $m$-dimensional state space, where the nonlinearity in propagation equations 
makes it difficult to track the state vector. Thus, it is quite difficult to forecast the
long-term behavior of an evolving SITR\ pattern for a T-bistable loop. Another
potential source of unpredictability is the interference by the pacer. If
reentry occurs, then the pacer and the reentrant circuit are thrown out of
phase so a phase resetting pacer pulse may unexpectedly end the reentrant circulation.

A number of prior studies of a different nature from ours also find complex SITR patterns in simple settings, such as in sheets [3, 25], loops [44] and aggregates [35] of animal cardiac tissue. In [3] and [25] complex patterns in monolayers of growing cell cultures are linked to changing densities and to local inhomogeneities, respectively. In [35] bursting beat patterns are studied in cell aggregates under external pacing. Closer to the subject matter of this paper, in [44] a ring-shaped lab preparation is considered with two localized pacemakers. Threshold relations for handling the pacers in this loop appear in the mathematical discussion. In a different class of studies, namely, the modulated parasystole, a permanent ectopic rhythm coexists with the sinus or another competing rhythm. In such relatively simple systems, complex beat patterns arise that are highly sensitive to parameter changes [23, 28, 43]. Like the TP threshold in our model, a mathematical relation with a discontinuity is responsible for the complexity of temporal patterns in parasystole; see [23, 43]. 

We omitted many features from our discussion here that one might include in an
enhanced version of this model. For example, UB may be treated functionally by 
refining the UB time window and considering bidirectional propagation in the loop, 
as noted above. Other enhancements to this model might include adding memory and 
latency to the APD restitution function. The inclusion of memory is especially important in clarifying the dynamic significance of delayed responses (not to mention improved fits to the APD restitution data). Additional directions for exploring the dynamics of the loop-pacer system include consideration of multiple exit/entry points into the loop that can affect reentry, a systematic approach to heterogeneity and also some theory to better model the time lag $\delta_{n}$ that influences the loop-pacer interactions. A somewhat different direction that may be explored would be to incorporate a programmed shock routine to artificially end long bursts of fast, reentrant beats.

With some of the above enhancements the SITR patterns will likely become more
complex and less predictable, though it is also possible that some of these
enhancements (e.g. memory) can have a moderating effect. Understanding the
precise mechanisms that generate such complex SITR patterns may in turn offer
new insights into the nature and causes of long-term temporal patterns of
tachyarrhythmia occurrences.

\medskip

\begin{center}
REFERENCES
\end{center}

\medskip

[1] Anastasiou-Nana, M.I., R.L. Menlove, J.N. Nanas and J.L. Anderson,
\textit{Changes in spontaneous variability of ventricular ectopic activity as
a function of time in patients with chronic arrhythmias}, Circulation, 78
(1988), 286-295.

[2] Banville, I. and R.A. Gray, \textit{Effects of action potential duration
and conduction velocity restitution on alternans and the stability of
arrhythmias}, J. Cardiovascular Electrophysiol., 13 (2002), 1141-1149.

[3] Bub, G., K. Tateno, A. Shrier and L. Glass, \textit{Spontaneous initiation
and termination of complex rhythms in cardiac cell culture}, J. Cardiovascular
Electrophysiol., 14 (2003), S229-S236.

[4] Cain, J.W. \textit{Criterion for stable reentry in a ring of cardiac
tissue}, J. Math. Biology, 55 (2007), 433--448.

[5] Cain, J.W. and D.G. Schaeffer, \textit{Two-term asymptotic approximation of
a cardiac restitution curve}, SIAM Rev., 48 (2006), 537-546.

[6] Cain, J.W., E.G. Tolkacheva, D.G. Schaeffer and D.J. Gauthier,
\textit{Rate-dependent propagation of cardiac action potentials in a
one-dimensional fiber}, Phys. Rev. E, 70 (2004), 061906.

[7] Cao, J-M., Z. Qu, Y-H. Kim, T-J. Wu, A. Garfinkel, J.N. Weiss, H.S.
Karagueuzian and P-S. Chen, \textit{Spatiotemporal heterogeneity in the
induction of ventricular fibrillation by rapid pacing: Importance of cardiac
restitution properties}, Circ. Research, 84 (1999), 1318-1331.

[8] Chen, X., F.H. Fenton and R.A. Gray, \textit{Head-tail interactions in
numerical simulations of reentry in a ring of cardiac tissue}, Heart Rhythm, 2
(2005), 851-859.

[9] S.S. Cheng, \textit{Partial Difference Equations}, CRC Press, Boca Raton, 2003.

[10] Cherry, E.M. and F.H. Fenton, \textit{Suppression of alternans and
conduction blocks despite steep APD\ restitution: Electrotonic, memory, and
conduction velocity restitution effects}, Am. J. Physiol. 286 (2004), H2332-H2341.

[11] Cytrynbaum, E. and J.P. Keener, \textit{Stability conditions for the
traveling pulse: Modifying the restitution hypothesis}, Chaos, 12 (2002), 788-799.

[12] Derksen, R., H.V.M. van Rijen, R. Wilders, S. Tasseron, R.N.W. Hauer,
W.L.C. Rutten and J.M.T. de Bakker, \textit{Tissue discontinuities affect conduction
velocity restitution: A mechanism by which structural barriers may promote
wave break}, Circulation, 108 (2003), 882-888.

[13] Chialvo, D.R., D.C. Michaels and J. Jalife, \textit{Supernormal
excitability as a mechanism of chaotic dynamics of activation in cardiac
Purkinje fibers,} Circ. Research, 66 (1990), 525-545.

[14] Courtemanche, M., L. Glass and J.P. Keener, \textit{Instabilities of a
propagating pulse in a ring of excitable media}, Phys. Rev. Lett., 70 (1993), 2182-2185.

[15] Courtemanche, M., J.P. Keener and L. Glass, \textit{A delay equation
representation of pulse circulation on a ring in excitable media}, SIAM J.
Appl. Math., 56 (1996), 119-142.

[16] Courtemanche, M. and A. Vinet, \textit{Reentry in excitable media}, in: A.
Beuter, L. Glass, M.C. Mackey and M.S. Titcombe (Ed.s), \textit{Nonlinear
Dynamics in Physiology and Medicine}, Chapter 7, Springer, New York, 2003.

[17] Elaydi, S.N. \textit{An Introduction to Difference Equations}, (2nd ed)
Springer, New York, 1999.

[18] Fox, J.J., E. Bodenschatz and R.F. Gilmour, \textit{Period-doubling
instability and memory in cardiac tissue}, Phys. Rev. Lett., 89 (2002), 138101(04).

[19] Fox, J.J., R.F. Gilmour and E. Bodenschatz, \textit{Conduction block in
one-dimensional heart fibers,} Phys. Rev. Lett., 89 (2002), 198101(04).

[20] Frame, L.H. and M.B. Simson, \textit{Oscillations of conduction, action
potential duration and refractoriness: A mechanism for spontaneous termination
of reentrant tachycardias}, Circulation, 78 (1988), 2182-2185.

[21] Gilmour, R.F., M.A. Watanabe and N.F. Otani, \textit{Restitution
properties and dynamics of reentry}, In: \textit{Cardiac Electrophysiology:
From Cell to Bedside}, 3rd ed., W.B. Saunders, London, 1999, 378-385.

[22] Girouard, S.D., J.M. Pastore, K.R. Laurita, K.W. Gregory and D.S.
Rosenbaum, \textit{Optical mapping in a new guinea pig model of ventricular
tachycardia reveals mechanisms for multiple wavelengths in a single reentrant
circuit}, Circulation, 93 (1996), 603-613.

[23] Glass, L., A.L. Goldberger, M. Courtemanche and A. Shrier,
\textit{Nonlinear dynamics, chaos and complex cardiac arrhythmias}, Proc. R.
Soc. Lond. A 413 (1987), 9-26.

[24] Hall, G.M., S. Bahar and D.J. Gauthier, \textit{Prevalence of
rate-dependent behaviors in cardiac muscle}, Phys. Rev. Lett., 82 (1999), 2995-2998.

[25] Hwang, S-M., T.Y. Kim and K.J. Lee, \textit{Complex-periodic spiral waves
in confluent cardiac cell cultures induced by localized inhomogeneities},
Proc. Nat. Acad. Sci. USA, 102 (2005), 10363-10368.

[26] Ito, H. and L. Glass, \textit{Theory of reentrant excitation in a ring of
cardiac tissue,} Physica D, 56 (1992), 84-106.

[27] Jack, J.J., D. Noble and R.W. Tsien, \textit{Electric Current Flow in
Excitable Cells}, Clarendon Press, Oxford, 1975.

[28] Jalife, J., C. Antzelevich and G.K. Moe, \textit{The case for modulated
parasystole}, Pace, 5 (1982), 911-926.

[29] Kalb, S.S., H.M. Dobrovolny, E.G. Tolkacheva, S.F. Idriss, W. Krassowska
and D.J. Gauthier, \textit{The restitution portrait: A new method for
investigating rate-dependent restitution}, J. Cardiovascular Electrophysiol.,
15 (2004), 698-709.

[30] Keener, J.P. \textit{Arrhythmias by dimension}, In: \textit{An
Introduction to Mathematical Modeling in Physiology, Cell Biology and
Immunology}, J. Sneyd (Ed.), 57-81, American Mathematical Society, 2002.

[31] Kocic, V.L. and G. Ladas, \textit{Global Behavior of Nonlinear Difference
Equations of Higher Order with Applications}, Kluwer, Dordrecht, 1993.

[32] Koller, M.L., M.L. Riccio and R.F. Gilmour, \textit{Dynamic restitution of action
potential duration during electrical alternans and ventricular fibrillation},
Am. J. Physiol. 275 (1998), H1635-H1642.

[33] Koller, M.L., S.K.G. Maier, A.R. Gelzer, W.R. Bauer, M. Meesman and R.F.
Gilmour, \textit{Altered dynamics of action potential restitution and alternans in
humans with structural heart disease}, Circulation, 112 (2005), 1542-1548.

[34] Krebs, M.E., J.M. Szwed, T. Shinn, W.M. Miles and D.P. Zipes,
\textit{Short-term rapid ventricular pacing prolongs ventricular
refractoriness in patients}, J. Cardiovascular Electrophysiol., 9 (1998), 1036-1042.

[35] Kunysz, A.M., A. Shrier and L. Glass, \textit{Bursting behavior during
fixed-delay stimulation of spontaneously beating chick heart cell aggregates},
Am. J. Physiol. 273 (1997), C331-C346.

[36] Lampert, R., L. Rosenfeld, W. Batsford, F. Lee and C. McPherson,
\textit{Circadian variation of sustained ventricular tachycardia in patients
with coronary artery disease and implantable cardioverter defibrillators},
Circulation, 90 (1994), 241-247.

[37] Liebovitch, L.S., A.T. Todorov, M. Zochowski, D. Scheurle, L. Colgin, M.A.
Wood, K.A. Ellenbogen, J.M. Herre and R.C. Bernstein, \textit{Nonlinear
properties of cardiac rhythm abnormalities}, Phys. Rev. E, 59 (1999), 3312-3319.

[38] Lown, B. \textit{Sudden cardiac death: The major challenge confronting
contemporary cardiology}, Am. J. Cardiol., 43 (1979), 313-328.

[39] Manios, E.G., E.M. Kallergis, E.M. Kanoupakis, H.E. Mavrakis, H.K.
Mouloudi, N.K. Klapsinos and P.E. Vardas, \textit{Effects of successful
cardioversion of persistent atrial fibrillation on right ventricular
refractoriness and repolarization}, Europace, 7 (2005), 34-39.

[40] Mickens, R. \textit{Difference Equations: Theory and Applications} (2nd
ed), CRC Press, Boca Raton, 1991.

[41] Mines, G.R. \textit{On dynamic equilibrium in the heart}, J. Physiol.
(London), 46 (1913), 349-383.

[42] Mitchell, C.C. and D.G. Schaeffer, \textit{A two-current model for the
dynamics of cardiac membrane,} Bull. Math. Biology, 65 (2003), 767-793.

[43] Moe, G.K., J. Jalife, W.J. Mueller and B. Moe, \textit{A mathematical
model of parasystole and its application to clinical arrhythmias},
Circulation, 56 (1977), 968-979.

[44] Nagai, Y., H. Gonzalez, A. Shrier and L. Glass, \textit{Paroxysmal
starting and stopping of circulating waves in excitable media}, Phys. Rev.
Lett., 18 (2000), 184248(4).

[45] Nolasco, J.B. and R.W. Dahlen, \textit{A graphic method for the study of
alternation in cardiac action potentials}, J. Appl. Physiol., 25 (1968), 191-196.

[46] Otani, N.F. and R.F. Gilmour, \textit{Memory models for the electrical
properties of local cardiac systems}, J. Theor. Biol., 187 (1997), 409-436.

[47] Plonsey, P. and R.C. Barr, \textit{Bioelectricity: A Quantitative Approach
(}2nd ed.), Kluwer Academic/Plenum Publ., NY, 2000.

[48] Qu, Z., J.N. Weiss and A. Garfinkel, \textit{Spatiotemporal chaos in a
simulated ring of cardiac cells}, Phys. Rev. Lett., 78 (1997), 1387-1390.

[49] Quan, W. and Y. Rudy, \textit{Unidirectional block and reentry of cardiac
excitation: A model study}, Circ. Research, 66 (1990), 367-382.

[50] Sedaghat, H., C.M. Kent and M.A. Wood, \textit{Criteria for the
convergence, oscillation and bistability of pulse circulation in a ring of
excitable media}, SIAM J. Appl. Math., 66 (2005), 573-590.

[51] Sedaghat, H. \textit{Nonlinear Difference Equations: Theory with
Applications to Social Science Models}, Kluwer Academic, Dordrecht, 2003.

[52] Stein, K.M., J.S. Borer, C. Hochreiter, and P. Kligfield, Fractal
clustering of ventricular ectopy and sudden death in mitral regurgitation, J.
Electrocardiol., (suppl) 25 (1992), 178-181.

[53] Stubna, M.D., R.H. Rand and R.F. Gilmour, \textit{Analysis of a nonlinear
partial difference equation, and its application to cardiac dynamics}, J.
Difference Equations and Appl., 8 (2002), 1147-1169.

[54] Ten Tusscher, K.H.W.J., O. Bernus, R. Hren and A.V. Panfilov,
\textit{Comparison of electrophysiological models for human ventricular cells
and tissues}, Prog. Biophys. Molecular Biology, 90 (2006), 326-345.

[55] Vinet, A. \textit{Quasiperiodic circus movement in a loop model of
cardiac tissue: Multistability and low dimensional equivalence}, Ann. Biomed.
Eng., 28 (2000), 704-720.

[56] Watanabe, M.A., F.H. Fenton, S.J. Evans, H.M. Hastings and A. Karma,
\textit{Mechanisms for discordant alternans}, J. Cardiovascular
Electrophysiol., 12 (2001), 196-206.

[57] Watanabe, M.A. and M.L. Koller, \textit{Mathematical analysis of dynamics
of cardiac memory and accommodation: Theory and experiment,} Am. J. Physiol.,
282 (2002), H1534-H1547.

[58] Winkle, R.A., D.C. Derrington, J.S. Schroeder, \textit{Characteristics of
ventricular tachycardia in ambulatory patients}, Am. J. Cardiol., 39 (1977), 487-492.

[59] Wood, M.A., P.M. Simpson, B.S. Stambler, J.M. Herre, R.C. Bernstein and
K.A. Ellenbogen, \textit{Long-term temporal patterns of ventricular
tachyarrhythmias}, Circulation, 91 (1995), 2371-2377.

[60] Wood, M.A., P.M. Simpson, W.B. London, B.S. Stambler, J.M. Herre, R.C.
Bernstein, K.A. Ellenbogen, \textit{Circadian patterns of ventricular
tachycardia in patients with implantable cardioverter defibrillators}, J. Am.
Coll. Cardiol., 25 (1995), 901-907.

[61] Wood, M.A., P.M. Simpson, L.S. Liebovitch, A.T. Todorov and K.A.
Ellenbogen, \textit{Temporal patterns of ventricular tachyarrhythmias:
Insights from the implantable cardioverter-defibrillator}, in: S.B. Dunbar,
K.A. Ellenbogen and A.E. Epstein (Ed.s), \textit{Sudden Cardiac Death: Past,
Present and Future}, Futura Pub. Co., Armonk, NY, 1997.

\end{document}